\documentclass[twoside, journal, onecolumn, double, 11pt]{IEEEtran}

\usepackage{epsfig}
\usepackage{graphicx}
\usepackage{amsmath}
\usepackage{amssymb}
\usepackage{float}
\usepackage{url}
\newtheorem{definition}{Definition}

\newtheorem{proposition}{Proposition}
\newtheorem{corollary}{Corollary}
\newtheorem{theorem}{Theorem}

\newtheorem{remark}{Remark}

\newtheorem{discussion}{Discussion}

\newcommand{\reals}{{\rm I\!R}}
\newcommand{\expectation}{{\rm I\!E}}

\begin{document}

\title{Tight Bounds for Symmetric Divergence Measures and a Refined Bound for Lossless Source Coding}

\markboth{Appears in the IEEE Transactions on Information Theory, February 2015}{I. SASON:
Tight Bounds for Symmetric Divergence Measures and a Refined Bound for Lossless Source Coding}

\author{Igal Sason\thanks{\em{I. Sason is  with the Department of Electrical Engineering, Technion--Israel
Institute of Technology, Haifa 32000, Israel (e-mail: sason@ee.technion.ac.il). This research work was supported
by the Israeli Science Foundation (ISF), grant number 12/12.}}}

\maketitle

\begin{abstract}
Tight bounds for several symmetric divergence measures are derived in terms of the total variation distance.
It is shown that each of these bounds is attained by a pair of 2 or 3-element probability distributions.
An application of these bounds for lossless source coding is provided, refining and improving a certain
bound by Csisz\'{a}r. Another application of these bounds has been recently introduced by
Yardi. {\em et al.} for channel-code detection.
\end{abstract}


\section{Introduction}
\label{section: Introduction}
Divergence measures are widely used in information theory, machine learning, statistics,
and other theoretical and applied branches of mathematics (see, e.g., \cite{Basseville},
\cite{CsiszarS_FnT}, \cite{Dragomir00}, \cite{ReidW11}).
The class of $f$-divergences, introduced independently in \cite{AliS, Csiszar67a} and
\cite{Morimoto63}, forms an important class of divergence measures. Their properties,
including relations to statistical tests and estimators, were studied, e.g., in
\cite{CsiszarS_FnT} and \cite{LieseV_IT2006}.

In \cite{Gilardoni06}, Gilardoni studied the problem of minimizing an arbitrary
{\em symmetric} $f$-divergence for a given total variation distance,
providing a closed-form solution of this optimization problem. In a follow-up paper
by the same author \cite{Gilardoni10},
Pinsker's and Vajda's type inequalities were studied for symmetric $f$-divergences,
and the issue of obtaining lower bounds on $f$-divergences for a fixed total
variation distance was further studied. One of the main results in \cite{Gilardoni10}
was a derivation of a simple closed-form lower bound on the relative entropy in
terms of the total variation distance, which suggests an improvement over
Pinsker's and Vajda's inequalities, and a derivation of a simple and reasonably
tight closed-form upper bound on the infimum of the relative entropy in terms of
the total variation distance.

An exact characterization of the minimum of the relative entropy subject to a fixed
total variation distance has been derived in \cite{FedotovHT_IT03} and \cite{Gilardoni06}.
More generally, sharp inequalities for $f$-divergences were recently studied in \cite{GSS_IT14}
as a problem of maximizing or minimizing an arbitrary $f$-divergence between two probability
measures subject to a finite number of inequality constraints on other $f$-divergences.
The main result stated in \cite{GSS_IT14} is that such infinite-dimensional
optimization problems are equivalent to optimization problems over finite-dimensional
spaces where the latter are numerically solvable.

Following previous work, {\em tight} bounds on symmetric $f$-divergences and related distances
are derived in this paper. An application of these bounds for lossless source coding is provided,
refining and improving a certain bound by Csisz\'{a}r from 1967 \cite{Csiszar67b}.

The paper is organized as follows: preliminary material is introduced in
Section~\ref{section: Preliminaries}, tight bounds for several symmetric divergence
measures, which are either symmetric $f$-divergences or related symmetric distances,
are derived in Section~\ref{section: Derivation of Tight Bounds on Symmetric Distance Measures};
these bounds are expressed in terms of the total variation distance, and their tightness is
demonstrated. One of these bounds is used in Section~\ref{section: bound for lossless source coding}
for the derivation of an improved and refined bound for lossless source coding.

\section{Preliminaries}
\label{section: Preliminaries}

We introduce, in the following, some preliminaries and notation that are
essential to this paper.

\begin{definition}
Let $P$ and $Q$ be two probability distributions with a
common $\sigma$-algebra $\mathcal{F}$.
The {\em total variation distance} between $P$ and $Q$ is defined by
\begin{align}
d_{\text{TV}}(P, Q)  \triangleq \sup_{A \in \mathcal{F}} |P(A) - Q(A)|.
\label{eq: total variation distance}
\end{align}
\label{definition: total variation distance}
\end{definition}

If $P$ and $Q$ are defined on a countable set, \eqref{eq: total variation distance}
is simplified to
\begin{equation}
d_{\text{TV}}(P, Q) = \frac{1}{2} \sum_{x} \bigl|P(x) - Q(x)\bigr| =
\frac{||P-Q||_1}{2}
\label{eq: the L1 distance is twice the total variation distance}
\end{equation}
so it is equal to one-half the $L_1$-distance between $P$ and $Q$.

\vspace*{0.2cm}
\begin{definition}
Let $f \colon (0, \infty) \rightarrow \reals$ be a convex function with $f(1)=0$,
and let $P$ and $Q$ be two probability distributions.
The {\em $f$-divergence} from $P$ to $Q$ is defined by
\begin{equation}
D_f(P||Q) \triangleq \sum_{x} Q(x) \, f\left(\frac{P(x)}{Q(x)}\right)
\label{eq:f-divergence}
\end{equation}
with the convention that
\begin{align*}
&\ 0 f\Bigl(\frac{0}{0}\Bigr) = 0, \quad
f(0) = \lim_{t \rightarrow 0^+} f(t), \\
&\ 0 f\Bigl(\frac{a}{0}\Bigr) = \lim_{t \rightarrow 0^+}
t f\Bigl(\frac{a}{t}\Bigr) = a \lim_{u \rightarrow \infty} \frac{f(u)}{u}, \quad \forall \, a > 0.
\end{align*}
\label{definition:f-divergence}
\end{definition}

\begin{definition}
An $f$-divergence is {\em symmetric} if $D_f(P||Q) = D_f(Q||P)$ for every $P$ and $Q$.
\end{definition}

\vspace*{0.1cm}
Symmetric $f$-divergences include (among others) the squared Hellinger distance where $$f(t) = (\sqrt{t}-1)^2, \quad
D_f(P||Q) = \sum_x \left(\sqrt{P(x)} - \sqrt{Q(x)}\right)^2,$$ and the total variation distance in
\eqref{eq: the L1 distance is twice the total variation distance} where $f(t) = \frac{1}{2} \, |t-1|.$

An $f$-divergence is symmetric if and only if the function $f$ satisfies the equality (see \cite[p.~765]{Gilardoni06})
\begin{equation}
f(u) = u \, f\left(\frac{1}{u}\right) + a (u-1), \quad \forall \, u \in (0, \infty)
\label{eq: characterization of f of a symmetric f-divergence}
\end{equation}
for some constant $a$. If $f$ is differentiable at $u=1$ then a differentiation of both sides of equality
\eqref{eq: characterization of f of a symmetric f-divergence} at $u=1$ gives that $a = 2 f'(1)$.

Note that the relative entropy (a.k.a. the Kullback-Leibler divergence)
$D(P||Q) \triangleq \sum_{x} P(x) \log\left(\frac{P(x)}{Q(x)}\right)$
is an $f$-divergence with $f(t) = t \log(t), \; t>0$; its dual, $D(Q||P)$, is an f-divergence with
$f(t) = -\log(t), \; t>0$; clearly, it is an asymmetric $f$-divergence since $D(P||Q) \neq D(Q||P)$ .

\vspace*{0.1cm}
The following result, which was derived by Gilardoni (see \cite{Gilardoni06, Gilardoni10}),
refers to the infimum of a symmetric $f$-divergence for a fixed value of the total variation distance:
\begin{theorem}
Let $f \colon (0, \infty) \rightarrow \reals$ be a convex function with $f(1)=0$,
and assume that $f$ is twice differentiable. Let
\begin{equation}
L_{D_f}(\varepsilon) \triangleq \inf_{P, Q \colon \, d_{\text{TV}}(P,Q) = \varepsilon} D_f(P||Q),
\quad \forall \, \varepsilon \in [0,1]
\label{eq:definition of infimum of f-divergence under a total variation distance}
\end{equation}
be the infimum of the $f$-divergence for a given total variation distance. If $D_f$
is a symmetric $f$-divergence, and $f$ is differentiable at $u=1$, then
\begin{equation}
L_{D_f}(\varepsilon) = (1-\varepsilon) \, f\left(\frac{1+\varepsilon}{1-\varepsilon}\right)
- 2 f'(1) \, \varepsilon, \quad \forall \, \varepsilon \in [0,1].
\label{eq: infimum of a symmetric f-divergence for a given total variation distance}
\end{equation}
\label{theorem: lower bound on symmetric f-divergence in terms of the total variation distance}
\end{theorem}

\vspace*{0.1cm}
Consider an arbitrary symmetric $f$-divergence.
Note that it follows from \eqref{eq: characterization of f of a symmetric f-divergence}
and \eqref{eq: infimum of a symmetric f-divergence for a given total variation distance}
that the infimum in \eqref{eq:definition of infimum of f-divergence under a total variation distance},
is attained by the pair of 2-element probability distributions where
$$P = \left( \frac{1-\varepsilon}{2}, \, \frac{1+\varepsilon}{2} \right), \quad
Q = \left( \frac{1+\varepsilon}{2}, \, \frac{1-\varepsilon}{2} \right)$$
(or by switching $P$ and $Q$ since $D_f$ is assumed to be a symmetric divergence).

Throughout this paper, the logarithms are on base~$e$ unless
the base of the logarithm is stated explicitly.

\section{Derivation of Tight Bounds on Symmetric Divergence Measures}
\label{section: Derivation of Tight Bounds on Symmetric Distance Measures}
The following section introduces tight bounds for several symmetric divergence
measures for a fixed value of the total variation distance. The statements are
introduced in
Section~\ref{subsection: Tight Bounds on the Bhattacharyya Distance}--\ref{subsection: Bounds on Jeffreys' divergence},
their proof are provided in \eqref{subsection: Proofs}, followed by discussions
on the statements in Section~\ref{subsection: Discussions on the Tight Bounds}.

\subsection{Tight Bounds on the Bhattacharyya Coefficient}
\label{subsection: Tight Bounds on the Bhattacharyya Distance}
\begin{definition}
Let  $P$ and $Q$ be two probability distributions that are defined on the same
set. The {\em Bhattacharyya coefficient} \cite{Kailath67} between $P$ and $Q$ is given by
\begin{align}
Z(P, Q) \triangleq \sum_{x} \sqrt{P(x) \, Q(x)} \, .
\label{eq: Bhattacharyya distance}
\end{align}
The {\em Bhttacharyya distance} is defined as minus the logarithm of the Bhattacharyya
coefficient, so that it is zero if and only if $P=Q$, and it is non-negative in general
(since $0 \leq Z(P,Q) \leq 1$, and $Z(P,Q)=1$ if and only if $P=Q$).
\label{definition: probability metrics}
\end{definition}

\begin{proposition}
Let $P$ and $Q$ be two probability distributions. Then, for a fixed value
$\varepsilon \in [0,1]$ of the total variation distance (i.e., if
$d_{\text{TV}}(P,Q)=\varepsilon$), the respective Bhattacharyya
coefficient satisfies the inequality
\begin{align}
1-\varepsilon \leq Z(P, Q) \leq \sqrt{1 - \varepsilon^2}.
\label{eq: tight bounds on the Bhattacharyya distance for a given total variation distance}
\end{align}
Both upper and lower bounds are tight: the upper bound is attained by
the pair of 2-element probability distributions
$$P = \left( \frac{1-\varepsilon}{2}, \, \frac{1+\varepsilon}{2} \right), \quad
Q = \left( \frac{1+\varepsilon}{2}, \, \frac{1-\varepsilon}{2} \right),$$ and the
lower bound is attained by the pair of 3-element probability distributions
$$P=(\varepsilon, 1-\varepsilon, 0), \quad Q = (0, 1-\varepsilon, \varepsilon).$$
\label{proposition: tight bounds on the Bhattacharyya distance for a given total variation distance}
\end{proposition}

\subsection{A Tight Bound on the Chernoff Information}
\label{subsection: A Tight Bound on the Chernoff Information}
\begin{definition}
The {\em Chernoff information} between two probability distributions $P$ and $Q$,
defined on the same set, is given by
\begin{align}
& C(P,Q) \triangleq -\min_{\lambda \in [0,1]} \; \log \left( \sum_{x}
P(x)^{\lambda} \, Q(x)^{1-\lambda} \right). \label{eq: Chernoff information}
\end{align}
\label{definition: Chernoff information and relative entropy}
\end{definition}

Note that
\begin{align}
C(P,Q) &= \max_{\lambda \in [0,1]} \left\{ -\log \left( \sum_{x}
P(x)^{\lambda} \, Q(x)^{1-\lambda} \right) \right\} \nonumber \\
&= \max_{\lambda \in (0,1)} \, \Bigl\{(1-\lambda) \, D_{\lambda}(P, Q) \Bigr\}
\label{eq: connection between the Chernoff information and Renyi divergence}
\end{align}
where $D_{\lambda}(P, Q)$ designates the R\'{e}nyi divergence of order
$\lambda$ \cite{vanErvenH12}.
The endpoints of the interval $[0, 1]$ are excluded in the second line
of \eqref{eq: connection between the Chernoff information and Renyi divergence}
since the Chernoff information is non-negative, and the logarithmic function
in the first line
of \eqref{eq: connection between the Chernoff information and Renyi divergence}
is equal to zero at both endpoints.

\begin{proposition}
Let
\begin{align}
C(\varepsilon) \triangleq \min_{P, Q \colon \, d_{\text{TV}}(P,Q) = \varepsilon} C(P, Q),
\quad \forall \, \varepsilon \in [0,1]
\label{eq: minimum of the Chernoff information for a fixed total variation distance}
\end{align}
be the minimum of the Chernoff information for a fixed value $\varepsilon \in [0,1]$ of the
total variation distance.
This minimum indeed exists, and it is equal to
\begin{equation}
C(\varepsilon) = \left\{\begin{array}{ll}
                 -\frac{1}{2} \, \log(1-\varepsilon^2)   & \mbox{if $\varepsilon \in [0,1)$} \\[0.1cm]
                 +\infty                                 & \mbox{if $\varepsilon = 1$.}
\end{array}
\right.
\label{eq: characterization of the minimum of the Chernoff information for a fixed total variation distance}
\end{equation}
For $\varepsilon \in [0,1)$, it is achieved by
the pair of 2-element probability distributions
$P = \left( \frac{1-\varepsilon}{2}, \, \frac{1+\varepsilon}{2} \right)$, and
$Q = \left( \frac{1+\varepsilon}{2}, \, \frac{1-\varepsilon}{2} \right)$.
\label{proposition: minimum of the Chernoff information for a fixed total variation distance}
\end{proposition}

\begin{corollary}
For any pair of probability distributions $P$ and $Q$,
\begin{equation}
C(P,Q) \geq -\frac{1}{2} \log \Bigl( 1- \bigl(d_{\text{TV}}(P,Q) \bigr)^2 \Bigr).
\label{eq: lower bound on the Chernoff information given a total variation distance}
\end{equation}
and this lower bound is tight for a given value of the total variation distance.
\label{corollary: tight lower bound on the Chernoff information for a given total variation distance}
\end{corollary}

\vspace*{0.2cm}
\begin{remark}[An Application of Corollary~\ref{corollary: tight lower bound on the Chernoff information for a given total variation distance}]
From Corollary~\ref{corollary: tight lower bound on the Chernoff information for a given total variation distance},
a lower bound on the total variation distance implies a lower bound on the Chernoff
information; consequently, it provides an upper bound on the best achievable Bayesian probability
of error for binary hypothesis testing (see, e.g., \cite[Theorem~11.9.1]{Cover_Thomas}).
This approach has been recently used in \cite{YardiKV14} to obtain a lower bound on the Chernoff
information for studying a communication problem that is related to channel-code detection via
the likelihood ratio test (the authors in \cite{YardiKV14} refer to our previously
un-published manuscript \cite{Sason12_arXiv}, where this corollary first appeared).
\end{remark}

\subsection{A Tight Bound on the Capacitory Discrimination}
\label{subsection: A Tight Bound on the Capacitory Discrimination}
\label{subsection: Bounds on the Capacitory Discrimination}
The capacitory discrimination (a.k.a. the Jensen-Shannon divergence) is defined as follows:
\begin{definition}
Let $P$ and $Q$ be two probability distributions.
The capacitory discrimination between $P$ and $Q$ is given by
\begin{equation}
\begin{split}
\overline{C}(P,Q) &\ \triangleq D\left(P \, || \,
\frac{P+Q}{2}\right)+D\left(Q \, || \, \frac{P+Q}{2}\right) \\
&\ = 2 \left[H\left(\frac{P+Q}{2}\right) - \frac{H(P)+H(Q)}{2} \right]
\label{eq: capacitory discrimination}
\end{split}
\end{equation}
where $H(P) \triangleq -\sum_x P(x) \, \log P(x)$.
\label{definition: capacitory discrimination}
\end{definition}
This divergence measure was studied in \cite{BrietH09}, \cite{BurbeaR82},
\cite{EndresS_IT03}, \cite{GSS_IT14}, \cite{Lin91} and \cite{Topsoe_IT00}.
Due to the parallelogram identity for relative entropy (see, e.g.,
\cite[Problem~3.20]{Csiszar_Korner}),
it follows that $\overline{C}(P,Q) = \min \bigl\{ D(P||R) + D(Q||R) \bigr\}$
where the minimization is taken w.r.t. all probability distributions $R$.

\begin{proposition}
For a given value $\varepsilon \in [0,1]$ of the total variation
distance, the minimum of the capacitory discrimination is equal to
\begin{equation}
\min_{P, Q \colon \, d_{\text{TV}}(P,Q) = \varepsilon}
\overline{C}(P,Q) = 2 \, d\left(\frac{1-\varepsilon}{2} \,
\big|\big| \, \frac{1}{2} \right)
\label{eq: a tight lower bound on the capacitory discrimination}
\end{equation}
and it is achieved by the 2-element probability distributions
$P = \left( \frac{1-\varepsilon}{2}, \, \frac{1+\varepsilon}{2} \right)$, and
$Q = \left( \frac{1+\varepsilon}{2}, \, \frac{1-\varepsilon}{2} \right)$. In
\eqref{eq: a tight lower bound on the capacitory discrimination},
\begin{equation}
d(p||q) \triangleq p \log\left(\frac{p}{q}\right)+
(1-p)\log\left(\frac{1-p}{1-q}\right), \quad p, q \in [0,1],
\label{eq: d function}
\end{equation}
with the convention that $0 \log 0 =0$, denotes the divergence (relative
entropy) between the two Bernoulli distributions with parameters $p$ and $q$.
\label{proposition: minimum of the capacitory discrimination for a fixed total variation distance}
\end{proposition}

\begin{remark}
The lower bound on the capacitory discrimination was obtained independently
by Bri{\"{e}}t and Harremo{\"{e}}s (see  \cite[Eq.~(18)]{BrietH09} for $\alpha=1$)
whose derivation was based on a different approach.
\end{remark}

\vspace*{0.1cm}
The following result provides a measure of the concavity of the entropy function:
\begin{corollary}
For arbitrary probability distributions $P$ and $Q$, the following inequality holds:
\begin{align*}
H\left(\frac{P+Q}{2}\right) - \frac{H(P)+H(Q)}{2}
\geq d\left(\frac{1-d_{\text{TV}}(P,Q)}{2} \, \big|\big| \, \frac{1}{2} \right)
\end{align*}
and this lower bound is tight for a given value of the total variation distance.
\label{corollary: tight lower bound on the capacitory discrimination in terms of the total variation distance}
\end{corollary}

\subsection{Tight Bounds on Jeffreys' divergence}
\label{subsection: Bounds on Jeffreys' divergence}
\begin{definition}
Let $P$ and $Q$ be two probability distributions. Jeffreys' divergence \cite{Jeffreys46}
is a symmetrized version of the relative entropy, which is defined as
\begin{equation}
J(P,Q) \triangleq \frac{D(P||Q) + D(Q||P)}{2}.
\label{eq: Jeffreys' divergence}
\end{equation}
\end{definition}
This forms a symmetric $f$-divergence where $J(P,Q) = D_f(P||Q)$ with
\begin{equation}
f(t) = \frac{(t-1) \log(t)}{2} \, , \quad t > 0,
\label{eq: Jeffreys' divergence as an f-divergence}
\end{equation}
which is a convex function on $(0, \infty)$, and $f(1)=0$.

\vspace*{0.1cm}
\begin{proposition}
\begin{align}
&\min_{P,Q \colon d_{\text{TV}}(P,Q) = \varepsilon} J(P,Q)
= \varepsilon \, \log\left(\frac{1+\varepsilon}{1-\varepsilon}\right),
\quad \forall \, \varepsilon \in [0, 1),
\label{eq: a tight lower bound on Jeffreys' divergence in terms of the total variation distance} \\
&\inf_{P,Q \colon D(P||Q) = \varepsilon} J(P,Q)
= \frac{\varepsilon}{2},
\quad \forall \, \varepsilon > 0,
\label{eq: a tight lower bound on Jeffreys' divergence in terms of the relative entropy}
\end{align}
and the two respective suprema are equal to $+\infty$. The minimum
of Jeffreys' divergence in
\eqref{eq: a tight lower bound on Jeffreys' divergence in terms of the total variation distance},
for a fixed value $\varepsilon$ of the total
variation distance, is achieved by the pair of 2-element probability distributions
$P = \left( \frac{1-\varepsilon}{2}, \, \frac{1+\varepsilon}{2} \right)$ and
$Q = \left( \frac{1+\varepsilon}{2}, \, \frac{1-\varepsilon}{2} \right)$.
\label{proposition: tight lower bound on Jeffreys' divergence in terms of the total variation distance}
\end{proposition}

\subsection{Proofs}
\label{subsection: Proofs}

\subsubsection{Proof of Proposition~\ref{proposition: tight bounds on the Bhattacharyya distance for a given total variation distance}}
From \eqref{eq: the L1 distance is twice the total variation distance},
\eqref{eq: Bhattacharyya distance} and the Cauchy-Schwartz inequality, we have
\begin{align*}
d_{\text{TV}}(P,Q)
& = \frac{1}{2} \, \sum_{x} \left| P(x) - Q(x) \right| \\[0.1cm]
& = \frac{1}{2} \, \sum_{x} \left| \sqrt{P(x)} - \sqrt{Q(x)} \, \right|
\left( \sqrt{P(x)} + \sqrt{Q(x)} \, \right) \\[0.1cm]
& \leq \frac{1}{2} \, \left( \sum_{x} \left( \sqrt{P(x)} - \sqrt{Q(x)} \,
\right)^2 \right)^{\frac{1}{2}} \left( \sum_{x} \left( \sqrt{P(x)} +
\sqrt{Q(x)} \, \right)^2 \right)^{\frac{1}{2}} \\[0.1cm]
& = \frac{1}{2} \bigl(2-2Z(P,Q)\bigr)^{\frac{1}{2}} \bigl(2+2Z(P,Q)\bigr)^{\frac{1}{2}} \\[0.1cm]
& = \left(1 - Z^2(P,Q)\right)^{\frac{1}{2}}
\end{align*}
which implies that $Z(P,Q) \leq \left(1 - d_{\text{TV}}^2(P,Q)\right)^{\frac{1}{2}}$.
This gives the upper bound on the Bhattacharyya coefficient in
\eqref{eq: tight bounds on the Bhattacharyya distance for a given total variation distance}.
For proving the lower bound, note that
\begin{align*}
Z(P,Q) &= 1 - \frac{1}{2} \sum_{x} \Bigl(\sqrt{P(x)} - \sqrt{Q(x)}\Bigr)^2 \\[0.1cm]
&= 1 - \frac{1}{2} \sum_{x} |P(x)-Q(x)| \left(\frac{|\sqrt{P(x)} - \sqrt{Q(x)}|}
{\sqrt{P(x)} + \sqrt{Q(x)}} \right) \\[0.1cm]
&\geq 1 - \frac{1}{2} \sum_{x} |P(x)-Q(x)| =  1-d_{\text{TV}}(P,Q).
\end{align*}
The tightness of the bounds on the Bhattacharyya coefficient in terms of the total
variation distance is proved in the following. For a fixed value of the total
variation distance $\varepsilon \in [0,1]$, let $P$ and $Q$ be the pair of
2-element probability distributions
$P = \left( \frac{1-\varepsilon}{2}, \, \frac{1+\varepsilon}{2} \right)$
and $Q = \left( \frac{1+\varepsilon}{2}, \, \frac{1-\varepsilon}{2} \right)$.
This gives $$d_{\text{TV}}(P,Q) = \varepsilon, \quad Z(P,Q) = \sqrt{1-\varepsilon^2}$$
so the upper bound is tight. Furthermore, for the
pair of 3-element probability distributions $P = (\varepsilon, 1-\varepsilon, 0)$ and
$Q = (0, 1-\varepsilon, \varepsilon)$, we have
$$d_{\text{TV}}(P,Q) = \varepsilon, \quad Z(P,Q) = 1-\varepsilon$$
so also the lower bound is tight.

\begin{remark}
The lower bound on the Bhattacharyya coefficient in
\eqref{eq: tight bounds on the Bhattacharyya distance for a given total variation distance}
dates back to Kraft \cite[Lemma~1]{Kraft55}, though its proof was simplified here.
\end{remark}

\begin{remark}
Both the Bhattacharyya distance and coefficient are functions of the Hellinger distance, so
a tight upper bound on the Bhattacharyya coefficient in terms of the total variation distance
can be also obtained from a tight upper bound on the Hellinger distance (see \cite[p.~117]{GSS_IT14}).

\end{remark}

\subsubsection{Proof of Proposition~\ref{proposition: minimum of the Chernoff information for a fixed total variation distance}}
\begin{eqnarray*}
&& C(P,Q) \stackrel{(\text{a})}{\geq} -\log \left( \sum_{x} \sqrt{P(x) \, Q(x)} \right) \\
&& \hspace*{1.5cm} \stackrel{(\text{b})}{=} -\log \, Z(P,Q) \\
&& \hspace*{1.5cm} \stackrel{(\text{c})}{\geq} -\frac{1}{2} \log
\Bigl( 1- \bigl(d_{\text{TV}}(P,Q) \bigr)^2 \Bigr)
\end{eqnarray*}
where inequality~(a) follows by selecting the possibly sub-optimal value of
$\lambda = \frac{1}{2}$ in \eqref{eq: Chernoff information}, equality~(b)
holds by definition (see \eqref{eq: Bhattacharyya distance}),
and inequality~(c) follows from the upper bound on the Bhattacharyya distance in
\eqref{eq: tight bounds on the Bhattacharyya distance for a given total variation distance}.
By the definition in
\eqref{eq: minimum of the Chernoff information for a fixed total variation distance}, it
follows that
\begin{equation}
C(\varepsilon) \geq -\frac{1}{2} \, \log(1-\varepsilon^2).
\label{eq: lower bound on the Chernoff information given epsilon}
\end{equation}
In order to show that \eqref{eq: lower bound on the Chernoff information given epsilon} provides
a tight lower bound for a fixed value of the total variation distance $(\varepsilon)$,
note that for the pair of 2-element probability distributions $P$ and $Q$ in
Proposition~\ref{proposition: minimum of the Chernoff information for a fixed total variation distance},
the Chernoff information in \eqref{eq: Chernoff information} is given by
\begin{align}
& C(P,Q) = -\min_{\lambda \in [0,1]} \log \left( \frac{1-\varepsilon}{2}
\left(\frac{1+\varepsilon}{1-\varepsilon}\right)^{\lambda} +
\frac{1+\varepsilon}{2}
\left(\frac{1-\varepsilon}{1+\varepsilon}\right)^{\lambda} \right).
\label{eq:Chernoff information for a pair of 2-element probability distributions}
\end{align}
A minimization of the function in
\eqref{eq:Chernoff information for a pair of 2-element probability distributions}
gives that $\lambda = \frac{1}{2}$, and
\begin{align*}
C(P,Q) = -\frac{1}{2} \, \log(1-\varepsilon^2),
\end{align*}
which implies that the lower bound in
\eqref{eq: lower bound on the Chernoff information given epsilon} is tight.

\subsubsection{Proof of Proposition~\ref{proposition: minimum of the capacitory discrimination for a fixed total variation distance}}

In \cite[p.~119]{GSS_IT14}, the capacitory discrimination
is expressed as an $f$-divergence where
\begin{equation}
f(x) = x \log x - (x+1) \log(1+x) + 2 \log 2, \quad x > 0
\label{eq: capacitory discrimination expressed as an f-divergence}
\end{equation}
is a convex function with $f(1)=0$. The combination of
\eqref{eq: infimum of a symmetric f-divergence for a given total variation distance}
and \eqref{eq: capacitory discrimination expressed as an f-divergence} implies that
\begin{align}
& \inf_{P, Q \colon \, d_{\text{TV}}(P,Q) = \varepsilon} \overline{C}(P,Q) \nonumber \\
&= (1-\varepsilon) \, f\left(\frac{1+\varepsilon}{1-\varepsilon}\right)
- 2 \varepsilon f'(1) \nonumber \\[0.1cm]
&= (1+\varepsilon) \, \log(1+\varepsilon) +
(1-\varepsilon) \, \log(1-\varepsilon) \nonumber \\[0.1cm]
&= 2 \left[ \log 2 - h\left(\frac{1-\varepsilon}{2} \right) \right]
= 2 \, d\left(\frac{1-\varepsilon}{2} \, \big|\big| \, \frac{1}{2} \right).
\label{eq: infimum of the capacitory discrimination}
\end{align}
The last equality holds since $d(p||\frac{1}{2}) = \log 2 - h(p)$ for $p \in [0,1]$
where $h$ denotes the binary entropy function.
Note that the infimum in \eqref{eq: infimum of the capacitory discrimination}
is a minimum since for the pair of 2-element probability distributions
$P = \left( \frac{1-\varepsilon}{2}, \, \frac{1+\varepsilon}{2} \right)$
and $Q = \left( \frac{1+\varepsilon}{2}, \, \frac{1-\varepsilon}{2} \right)$,
we have
$$D\left(P \, || \, \frac{P+Q}{2}\right) = D\left(Q \, || \, \frac{P+Q}{2}\right)
= d\left(\frac{1-\varepsilon}{2} \, \big|\big| \, \frac{1}{2} \right),$$
so, $\overline{C}(P,Q) = 2 d\left(\frac{1-\varepsilon}{2} \, \big|\big| \, \frac{1}{2} \right)$.

\subsubsection{Proof of Proposition~\ref{proposition: tight lower bound on Jeffreys' divergence in terms of the total variation distance}}
Jeffreys' divergence is a symmetric $f$-divergence where the convex
function $f$ in \eqref{eq: Jeffreys' divergence as an f-divergence}
satisfies the equality $f(t) = t f(\frac{1}{t})$ for every $t > 0$
with $f(1)=0$. From
Theorem~\ref{theorem: lower bound on symmetric f-divergence in terms of the total variation distance},
it follows that
\begin{align*}
&\inf_{P,Q \colon d_{\text{TV}}(P,Q) = \varepsilon} J(P,Q)
= \varepsilon \, \log\left(\frac{1+\varepsilon}{1-\varepsilon}\right),
\quad \forall \, \varepsilon \in [0, 1).
\end{align*}
This infimum is achieved by the pair of 2-element probability distributions
$P = \left(\frac{1+\varepsilon}{2}, \frac{1-\varepsilon}{2}\right)$
and $Q = \left(\frac{1-\varepsilon}{2}, \frac{1+\varepsilon}{2}\right)$,
so it is a minimum. This proves
\eqref{eq: a tight lower bound on Jeffreys' divergence in terms of the total variation distance}.

Eq.~\eqref{eq: a tight lower bound on Jeffreys' divergence in terms of the relative entropy}
follows from \eqref{eq: Jeffreys' divergence} and the fact that, given the value of the relative
entropy $D(P||Q)$, its dual $(D(Q||P))$ can be made arbitrarily small.

The two respective suprema are equal to infinity because given the value of
the total variation distance or the relative entropy, the dual of the relative
entropy can be made arbitrarily large.

\subsection{Discussions on the Tight Bounds}
\label{subsection: Discussions on the Tight Bounds}

\begin{discussion}
Let
\begin{equation}
L(\varepsilon) \triangleq \inf_{P, Q \colon \, d_{\text{TV}}(P,Q) = \varepsilon} D(P||Q).
\label{eq: infimum of the relative entropy for a given total variation}
\end{equation}
The exact parametric equation of the
curve $(\varepsilon, L(\varepsilon))_{0<\varepsilon<1}$ was introduced in different forms
in \cite[Eq.~(3)]{FedotovHT_IT03}, \cite{Gilardoni06}, and \cite[Eq.~(59)]{ReidW11}.
For $\varepsilon \in [0,1)$, this infimum is attained by a pair of 2-element probability distributions
(see \cite{FedotovHT_IT03}). Due to the factor of one-half in the total variation distance of
\eqref{eq: the L1 distance is twice the total variation distance}, it follows that
\begin{align}
L(\varepsilon) = \min_{\beta \in [\varepsilon-1, \, 1-\varepsilon]} \left\{
\left(\frac{\varepsilon+1-\beta}{2}\right) \, \log \left(\frac{\beta-1-\varepsilon}{\beta-1+\varepsilon}\right)
+ \left(\frac{\beta+1-\varepsilon}{2}\right) \, \log\left(\frac{\beta+1-\varepsilon}{\beta+1+\varepsilon}\right) \right\}
\label{eq:Reid and Willimason's bound for the relative entropy}
\end{align}
where, it can be verified that the numerical minimization w.r.t. $\beta$ in
\eqref{eq:Reid and Willimason's bound for the relative entropy} can be restricted
to the interval $[\varepsilon-1, \, 0]$.

Since $C(P, Q) \leq \min \bigl\{D(P||Q), D(Q||P)\}$ (see \cite[Section~11.9]{Cover_Thomas}),
it follows from \eqref{eq: minimum of the Chernoff information for a fixed total variation distance} and
\eqref{eq: infimum of the relative entropy for a given total variation} that
\begin{equation}
C(\varepsilon) \leq L(\varepsilon), \quad \forall \, \varepsilon \in [0,1)
\label{eq: inequality for C and L}
\end{equation}
where the right and left-hand sides of \eqref{eq: inequality for C and L} correspond to the minima of the
relative entropy and Chernoff information, respectively, for a fixed value of the total variation
distance $(\varepsilon)$.
\begin{figure}[here!]
\begin{center}
\epsfig{file=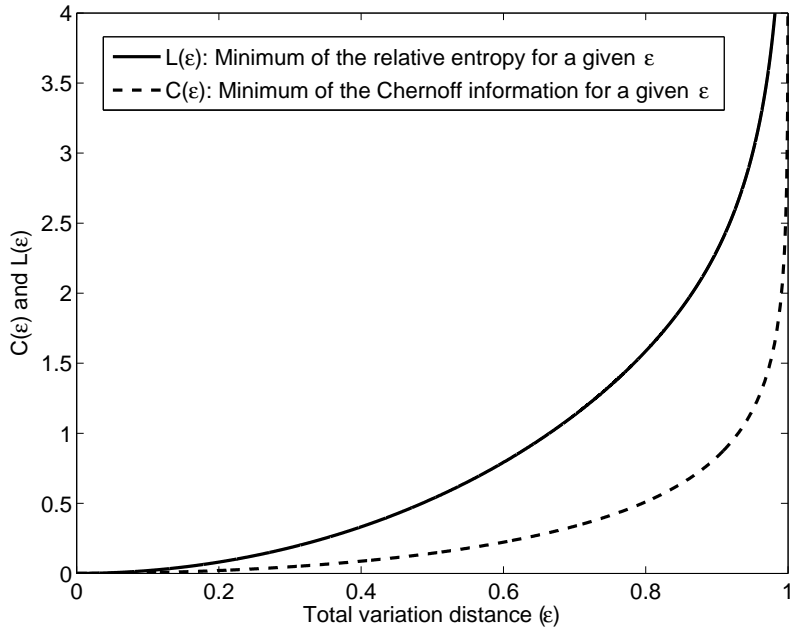,scale=0.6}
\caption{\label{Figure:plot of C versus L}
A plot of the minima of the Chernoff information and the relative entropy for
a given total variation distance $\varepsilon \in [0,1]$, denoted by $C(\varepsilon)$
and $L(\varepsilon)$, respectively; $C$ and $L$ are provided, respectively,
in Proposition~\ref{proposition: minimum of the Chernoff information for a fixed total variation distance}
and \cite[Theorem~2]{FedotovHT_IT03} or \cite[Eq.~(59)]{ReidW11} (see
\eqref{eq:Reid and Willimason's bound for the relative entropy}).}
\end{center}
\end{figure}
Figure~\ref{Figure:plot of C versus L} plots these minima
as a function of the total variation distance. For small values of $\varepsilon$,
$C(\varepsilon)$ and $L(\varepsilon)$, respectively, are approximately equal to $\frac{\varepsilon^2}{2}$
and $2 \varepsilon^2$ (note that Pinsker's inequality is tight for $\varepsilon \ll 1$), so
$$\lim_{\varepsilon \rightarrow 0} \frac{L(\varepsilon)}{C(\varepsilon)} = 4.$$
\end{discussion}

\vspace*{0.2cm}
\begin{discussion}
The lower bound on the capacitory discrimination in \eqref{eq: a tight lower bound on the capacitory discrimination},
expressed in terms of the total variation distance, forms a closed-form
expression of the bound by Tops{\o}e in \cite[Theorem~5]{Topsoe_IT00}.
The bound in \cite{Topsoe_IT00} is
\begin{equation}
\overline{C}(P,Q) \geq \sum_{\nu=1}^{\infty} \frac{\bigl(d_{\text{TV}}(P,Q)\bigr)^{2\nu}}{\nu(2\nu-1)}.
\label{eq:Topsoe}
\end{equation}
The equivalence of \eqref{eq: a tight lower bound on the capacitory discrimination} and \eqref{eq:Topsoe}
follows from the power series expansion of the binary entropy function
$$h(x) = \log 2 - \sum_{\nu=1}^{\infty} \frac{(1-2x)^{2\nu}}{2\nu (2\nu-1)}, \quad \forall \, x \in [0,1]$$
which yields that
\begin{align*}
\sum_{\nu=1}^{\infty} \frac{\bigl(d_{\text{TV}}(P,Q)\bigr)^{2\nu}}{\nu(2\nu-1)}
&= 2 \left[\log 2 - h\left(\frac{1-d_{\text{TV}}(P,Q)}{2}\right) \right] \\
&= 2 d\left(\frac{1-d_{\text{TV}}(P,Q)}{2} \,
\big|\big| \, \frac{1}{2}\right)
\end{align*}
where $d(\cdot || \cdot)$ is defined in \eqref{eq: d function}.
Note, however, that the proof here is more simple than the proof of \cite[Theorem~5]{Topsoe_IT00}
(which relies on properties of the triangular discrimination in \cite{Topsoe_IT00} and
previous theorems of this paper), and it also leads directly to a closed-form expression of
this bound. Consequently, one concludes that the lower bound in
\cite[Theorem~5]{Topsoe_IT00} is a special case of
Theorem~\ref{theorem: lower bound on symmetric f-divergence in terms of the total variation distance}
(see \cite{Gilardoni06} and \cite[Corollary~5.4]{GSS_IT14}), which provides a
lower bound on a symmetric $f$-divergence in terms of the total variation
distance.
\end{discussion}

\section{A Bound for Lossless Source Coding}
\label{section: bound for lossless source coding}

We illustrate in the following a use of
Proposition~\ref{proposition: tight lower bound on Jeffreys' divergence in terms of the total variation distance}
for the derivation of an improved and refined bound for
lossless source coding. This tightens, and also refines
under a certain condition, a bound by Csisz\'{a}r \cite{Csiszar67b}.

Consider a memoryless and stationary source with alphabet
$\mathcal{U}$ that emits symbols according to a probability
distribution $P$, and assume that a uniquely decodable (UD) code
with an alphabet of size $d$ is used. It is well known that such a
UD code achieves the entropy of the source if and only if
the length $l(u)$ of the codeword that is assigned to each
symbol $u \in \mathcal{U}$ satisfies the equality
$$l(u) = -\log_d P(u), \quad \forall \, u \in \mathcal{U}.$$
This corresponds to a dyadic source where, for every
$u \in \mathcal{U}$, we have $P(u) = d^{-n_u}$
with a natural number $n_u$; in this case, $l(u) = n_u$
for every symbol $u \in \mathcal{U}$. Let
$\overline{L} \triangleq \expectation[L]$ designate the
average length of the codewords, and
$H_d(U) \triangleq -\sum_{u \in \mathcal{U}} P(u) \, \log_d P(u)$
be the entropy of the source (to the base $d$). Furthermore, let
$c_{d,l} \triangleq \sum_{u \in \mathcal{U}} d^{-l(u)}.$
According to the Kraft-McMillian inequality (see
\cite[Theorem~5.5.1]{Cover_Thomas}), the inequality
$c_{d,l} \leq 1$ holds in general for UD codes,
and the equality $c_{d,l}=1$ holds if
the code achieves the entropy of the source (i.e.,
$\overline{L} = H_d(U)$).

Define a probability distribution $Q_{d,l}$ by
\begin{equation}
Q_{d,l}(u) \triangleq \left(\frac{1}{c_{d,l}} \right) \, d^{-l(u)},
\quad \forall \, u \in \mathcal{U}
\label{eq:probability distribution Q}
\end{equation}
and let $\Delta_d \triangleq \overline{L} - H_d(U)$ designate the
average redundancy of the code. Note that for a UD code that achieves
the entropy of the source, its probability distribution $P$ is
equal to $Q_{d,l}$ (since $c_{d,l}=1$, and $P(u) = d^{-l(u)}$
for every $u \in \mathcal{U}$).

In \cite{Csiszar67b}, a generalization for UD source codes has been
studied by a derivation of an upper bound on the $L_1$ norm between
the two probability distributions $P$ and $Q_{d,l}$ as a function
of the average redundancy $\Delta_d$ of the code. To this end, straightforward
calculation shows that the relative entropy from $P$ to $Q_{d,l}$ is given by
\begin{equation}
D(P || Q_{d,l}) = \Delta_d \, \log d + \log \bigl(c_{d,l}\bigr).
\label{eq:relative entropy from P to Q}
\end{equation}
The interest in \cite{Csiszar67b} is in getting an upper bound that
only depends on the average redundancy $\Delta_d$ of the code, but
is independent of the distribution of the lengths of the codewords.
Hence, since the Kraft-McMillian inequality states that $c_{d,l} \leq 1$
for general UD codes, it is concluded in \cite{Csiszar67b} that
\begin{equation}
D(P || Q_{d,l}) \leq \Delta_d \, \log d.
\label{eq:upper bound on the relative entropy}
\end{equation}
Consequently, it follows from Pinsker's inequality that
\begin{equation}
\sum_{u \in \mathcal{U}} \bigl| P(u) - Q_{d,l}(u) \bigr| \leq \min
\bigl\{ \sqrt{2 \Delta_d \log d}, \, 2 \bigr\}
\label{eq:Csiszar's bound for lossless source coding}
\end{equation}
since also, from the triangle inequality, the sum on the left-hand side of
\eqref{eq:Csiszar's bound for lossless source coding} cannot exceed~2.
This inequality is indeed consistent with the fact that the probability
distributions $P$ and $Q_{d,l}$ coincide when $\Delta_d = 0$ (i.e., for
a UD code that achieves the entropy of the source).

At this point we deviate from the analysis in \cite{Csiszar67b}.
One possible improvement of the bound in \eqref{eq:Csiszar's bound for lossless source coding}
follows by replacing Pinsker's inequality with the result in
\cite{FedotovHT_IT03}, i.e., by taking into account the exact
parametrization of the infimum of the relative entropy for a
given total variation distance. This gives the following tightened bound:
\begin{equation}
\sum_{u \in \mathcal{U}} \bigl| P(u) - Q_{d,l}(u) \bigr| \leq 2 \;
L^{-1}(\Delta_d \log d)
\label{eq:first tightening of Csiszar's bound for lossless source coding}
\end{equation}
where $L^{-1}$ is the inverse function of $L$ in
\eqref{eq:Reid and Willimason's bound for the relative entropy}
(it is calculated numerically).

In the following, the utility of
Proposition~\ref{proposition: tight lower bound on Jeffreys' divergence in terms of the total variation distance}
is shown by refining the latter bound in
\eqref{eq:first tightening of Csiszar's bound for lossless source coding}. Let
$$\delta(u) \triangleq l(u) + \log_d P(u), \quad \forall \, u \in \mathcal{U}.$$

Calculation of the dual divergence gives
\begin{align}
&D(Q_{d,l} || P) \nonumber \\
&= \log d \, \sum_{u \in \mathcal{U}} Q_{d,l}(u) \log_d \left(\frac{Q_{d,l}(u)}{P(u)} \right) \nonumber \\
&= \log d \, \left[ -\frac{\log_d(c_{d,l})}{c_{d,l}} \sum_{u \in \mathcal{U}} d^{-l(u)}
- \frac{1}{c_{d,l}} \sum_{u \in \mathcal{U}} l(u) d^{-l(u)}
-\frac{1}{c_{d,l}} \sum_{u \in \mathcal{U}} \log_d P(u) \; d^{-l(u)} \right] \nonumber \\
&= -\log(c_{d,l}) - \frac{\log d}{c_{d,l}} \sum_{u \in \mathcal{U}} \delta(u) \, d^{-l(u)} \nonumber \\
&= -\log \bigl(c_{d,l}\bigr) - \frac{\log d}{c_{d,l}} \,
\sum_{u \in \mathcal{U}} P(u) \, \delta(u) \, d^{-\delta(u)} \nonumber \\
&= -\log \bigl(c_{d,l}\bigr) - \left(\frac{\log d}{c_{d,l}} \right)
\expectation \bigl[\delta(U) \, d^{-\delta(U)}\bigr]
\label{eq:relative entropy from Q to P}
\end{align}
and the combination of \eqref{eq: Jeffreys' divergence}, \eqref{eq:relative entropy from P to Q}
and \eqref{eq:relative entropy from Q to P} yields that
\begin{equation}
J(P, Q_{d,l}) = \frac{1}{2} \left[\Delta_d \log d - \left(\frac{\log d}{c_{d,l}}\right)
\expectation \bigl[\delta(U) \, d^{-\delta(U)}\bigr] \right].
\label{eq:Jeffreys' divergence between P and Q}
\end{equation}
In the continuation of this analysis, we restrict our attention to UD codes
that satisfy the condition
\begin{equation}
l(u) \geq \left\lceil \log_d \frac{1}{P(u)} \right\rceil, \quad \forall \, u \in \mathcal{U}.
\label{eq:condition for further sarpening the bound for lossless source coding}
\end{equation}
In general, it excludes Huffman codes; nevertheless, it is
satisfied by some other important UD codes such as the Shannon
code, Shannon-Fano-Elias code, and arithmetic coding (see, e.g.,
\cite[Chapter~5]{Cover_Thomas}). Since
\eqref{eq:condition for further sarpening the bound for lossless source coding}
is equivalent to the condition that $\delta$ is non-negative on
$\mathcal{U}$, it follows from
\eqref{eq:Jeffreys' divergence between P and Q} that
\begin{equation}
J(P, Q_{d,l}) \leq \frac{\Delta_d \log d}{2}
\label{eq:upper bound on Jeffreys' divergence}
\end{equation}
so, the upper bound on Jeffreys' divergence in
\eqref{eq:upper bound on Jeffreys' divergence} is
twice smaller than the upper bound on the relative
entropy in \eqref{eq:upper bound on the relative entropy}.
It is partially because the term $\log c_{d,l}$ is canceled
out along the derivation of the bound in
\eqref{eq:upper bound on Jeffreys' divergence},
in contrast to the derivation of the bound in
\eqref{eq:upper bound on the relative entropy}
where this term was upper bounded by zero (hence,
it has been removed from the bound) in order
to avoid its dependence on the length of the codeword
for each individual symbol.

\begin{figure}[here!]
\begin{center}
\epsfig{file=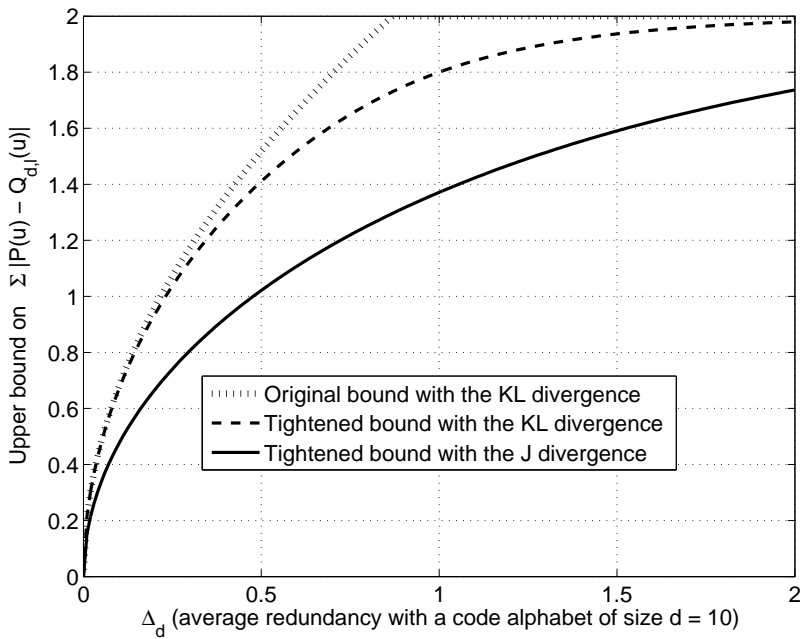,scale=0.6}
\caption{\label{Figure: comparison of bounds for lossless source coding}
Upper bounds on $\sum |P(u) - Q_{d,l}(u)|$ as a function of the average
redundancy $\Delta_d \triangleq \expectation[L]-H_d$ for a UD code with
an alphabet of size $d=10$. The original bound in
\eqref{eq:Csiszar's bound for lossless source coding} appears in \cite{Csiszar67b},
and the tightened bound that relies on the Kullback-Leibler (KL) divergence is
given in \eqref{eq:first tightening of Csiszar's bound for lossless source coding}.
The further tightening of this bound is restricted in this plot to UD codes
whose codewords satisfy the condition in
\eqref{eq:condition for further sarpening the bound for lossless source coding}.
The latter bound relies on
Proposition~\ref{proposition: tight lower bound on Jeffreys' divergence in terms of the total variation distance}
for Jeffreys' (J) divergence, and it is given in \eqref{eq:second tightening of Csiszar's bound for lossless source coding}.}
\end{center}
\end{figure}

Following Proposition~\ref{proposition: tight lower bound on Jeffreys' divergence in terms of the total variation distance},
for $x \geq 0$, let $\varepsilon \triangleq \varepsilon(x)$
be the unique solution in the interval $[0, 1)$ of the equation
\begin{equation}
\varepsilon \, \log\left(\frac{1+\varepsilon}{1-\varepsilon}\right) = x.
\label{eq:equation for J divergence}
\end{equation}
The combination of \eqref{eq: a tight lower bound on Jeffreys' divergence in terms of the total variation distance}
and \eqref{eq:upper bound on Jeffreys' divergence} implies that
\begin{equation}
\sum_{u \in \mathcal{U}} \bigl| P(u) - Q_{d,l}(u) \bigr|
\leq 2 \; \varepsilon \left( \frac{\Delta_d \log d}{2} \right).
\label{eq:second tightening of Csiszar's bound for lossless source coding}
\end{equation}
The bounds in \eqref{eq:Csiszar's bound for lossless source coding},
\eqref{eq:first tightening of Csiszar's bound for lossless source coding}
and \eqref{eq:second tightening of Csiszar's bound for lossless source coding}
are depicted in Figure~\ref{Figure: comparison of bounds for lossless source coding}
for UD codes where the size of their alphabet is $d=10$.

In the following, the bounds in
\eqref{eq:first tightening of Csiszar's bound for lossless source coding}
and \eqref{eq:second tightening of Csiszar's bound for lossless source coding}
are compared analytically for the case where the average redundancy is small
(i.e., $\Delta_d \approx 0$). Under this approximation, the bound in
\eqref{eq:Csiszar's bound for lossless source coding} (i.e., the original bound
from \cite{Csiszar67b}) coincides with its tightened version in
\eqref{eq:first tightening of Csiszar's bound for lossless source coding}.
On the other hand, since for $\varepsilon \approx 0$, the left-hand side of
\eqref{eq:equation for J divergence} is approximately $2 \varepsilon^2$,
it follows from \eqref{eq:equation for J divergence} that, for $x \approx 0$,
we have $\varepsilon(x) \approx \sqrt{\frac{x}{2}}.$
It follows that, if $\Delta_d \approx 0$, inequality
\eqref{eq:second tightening of Csiszar's bound for lossless source coding} gets
approximately the form
$$\sum_{u \in \mathcal{U}} \bigl| P(u) - Q_{d,l}(u) \bigr| \leq \sqrt{\Delta_d \log d}.$$
Hence, even for a small average redundancy, the bound in
\eqref{eq:second tightening of Csiszar's bound for lossless source coding}
improves \eqref{eq:Csiszar's bound for lossless source coding} by a factor of $\sqrt{2}$.
This conclusion is consistent with the plot in
Figure~\ref{Figure: comparison of bounds for lossless source coding}.

\section*{Acknowledgment}
The author thanks the anonymous reviewers for their helpful comments.

\end{document}